# Automated Ensemble-Based Segmentation of Adult Brain Tumors: A Novel Approach Using the BraTS AFRICA Challenge Data

Chiranjeewee Prasad Koirala[1*], Sovesh Mohapatra[1,2*], Advait Gosai[1], Gottfried Schlaug[2,3,4#]

[1]Manning College of Information and Computer Sciences, University of Massachusetts Amherst, MA
[2]Instititute for Applied Life Sciences, University of Massachusetts Amherst, MA
[3]Department of Biomedical Engineering, University of Massachusetts Amherst, MA
[4]Department of Neurology, Baystate Medical Center, and UMass Chan Medical School - Baystate Campus, Springfield, MA.

[*]contributed equally
[#]corresponding author

Corresponding author: E-mail: gschlaug@umass.edu;

## Abstract:

Brain tumors, particularly glioblastoma, continue to challenge medical diagnostics and treatments globally. This paper explores the application of deep learning to multi-modality magnetic resonance imaging (MRI) data for enhanced brain tumor segmentation precision in the Sub-Saharan Africa patient population. We introduce an ensemble method that comprises eleven unique variations based on three core architectures (UNet3D, ONet3D, SphereNet3D) and modified loss functions. The study emphasizes the need for both age- and population-based segmentation models, to fully account for the brain's complexities. Our findings reveal that the ensemble approach, combining different architectures, outperforms single models, leading to improved evaluation metrics. Specifically, the results exhibit Dice scores of 0.82, 0.82, and 0.87 for enhancing tumor, tumor core, and whole tumor labels respectively. These results underline the potential of tailored deep learning techniques in precisely segmenting brain tumors and lay groundwork for future work to fine-tune models and assess performance across different brain regions.

# 1. Introduction:

Brain cancer is among the deadliest types of cancers. Approximately 80% of individuals diagnosed with the malignant type of glioblastoma die within two years. This is in contrast to 90% of individuals affected with breast or prostate cancer who are expected to survive for a minimum of five years [1,2]. While high-income countries, particularly in the Global North, have witnessed an increase in cancer survival rates of around 10-30% over the past decades, this improvement has not translated to low and middle-income countries (LMICs). In fact, the African continent, in particular, has experienced the opposite trend, with the lowest and rapidly decreasing survival rates [3].

In the past decade, the increase in the use of deep learning and multi-modality MRI data has shown significant promise in precisely identifying and segmenting the brain tumors and its spread [4–6]. Along with age, it is also observed that different populations have different complexities in the brain [7,8]. Due to this, it is essential for not only creating age-based segmentation models but also population based models which allow for more precision. This situation also illustrates the fact that, owing to the various complexities present in the brain, a single architecture or model is insufficient to address all intricacies [9].

In this paper, we are focusing on BraTS Africa Challenge data which constitutes four different modalities (native T1, post-contrast T1-weighted (T1Gd), T2-weighted (T2), and T2 Fluid Attenuated Inversion Recovery (T2-FLAIR))[3]. Here, we introduce a novel ensemble method which allows us to create a tailored model for the Sub-African population.

## 2. Methodology

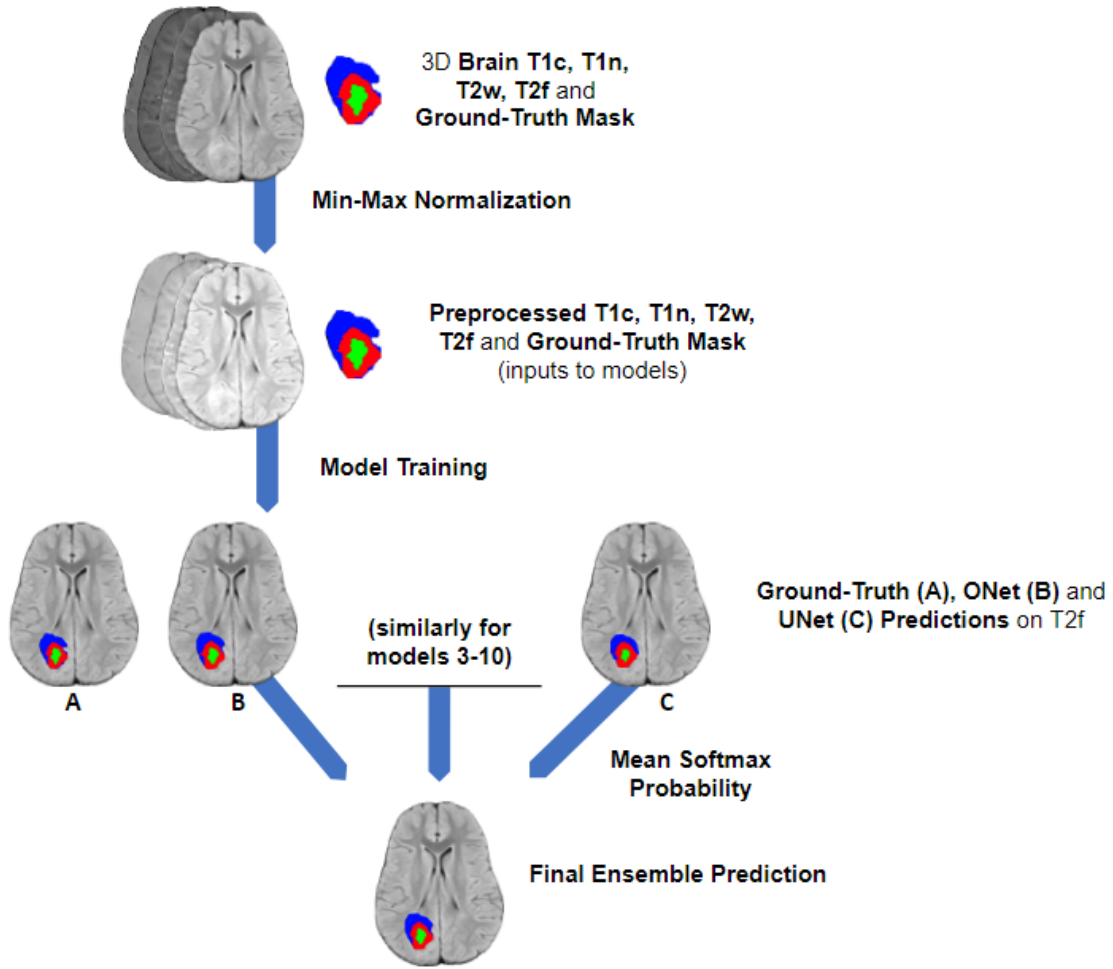

Figure - 1. Comprehensive workflow – visualizing the end-to-end process with model training and ensembling approaches

### 2.1. Model Architectures

In this work, we implemented eleven unique variations using three fundamental base architectures, each paired with different loss functions (further explained in section 2.2) and tailored hyperparameters. Figure - 2 illustrates the architecture for one of the variants of the UNet3D configurations.

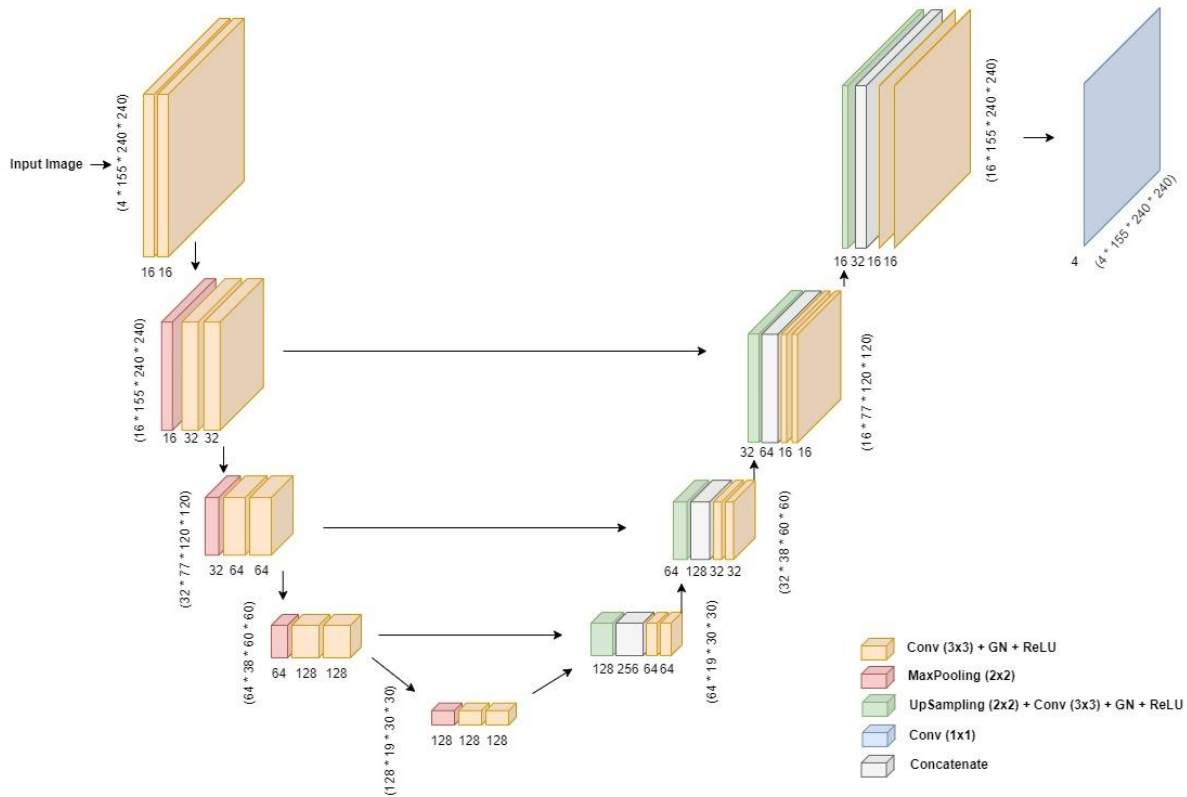

Figure - 2. Core UNet3D Architecture

### 2.1.1. UNet3D Family

Below are the different variations applied to the fundamental base architecture of UNet3D model:

1. **3D UNet**: UNet3d(in_channels = 4, n_classes = 4, n_channels = 16).
2. **Modified 3D UNet**: Modified3dUNet(in_channels = 4, n_classes = 4, n_channels = 16)[10].
3. **3D UNet with Deep Supervision**: UNet3dDeepSupervision(in_channels = 4, n_classes = 4, n_channels = 16), builds upon the conventional UNet base architecture. Within this architecture, segmentation masks are derived from both the third-to-last and second-to-last decoding layers. Subsequently, loss functions are applied to these intermediate outcomes and incorporated into the training phase. This integration introduces an element of deep supervision into the network's learning paradigm, enhancing its functionality.
4. **3D UNet GELU**: UNet3dGELU(in_channels=4, n_classes=4, n_channels=16). 3DUNet with all ReLU replaced by GELU activation function.
5. **3D UNet SingleConv**: UNet3dSingleConv(in_channels=4, n_classes=4, n_channels=16). 3DUNet with single convolution instead of double convolution per upscale/downscale.

6. 3DUnet Attention: UNet3dAtten(in_channels=4, n_classes=4, n_channels=16). 3DUNet with an attention layer.

### 2.1.2. ONet3D Family

Below are the different variations applied to the fundamental base architecture of ONet3D where the encoder-decoder sections are concatenated before the output convolution layer:

1. **3D ONet SingleConv Kernel_1**: ONet3DSingleConv1(in_channels=4, n_classes=4, n_channels=16, kernel_size = 1)
2. **3D ONet SingleConv Kernel_3**: ONet3DSingleConv3(in_channels=4, n_classes=4, n_channels=16, kernel_size = 3)
3. **3D ONet SingleConv Kernel_5**: ONet3DSingleConv5(in_channels=4, n_classes=4, n_channels=16, kernel_size = 5)
4. **3D ONet DoubleConv Kernel_1**: ONet3DDoubleConv1(in_channels=4, n_classes=4, n_channels=16, kernel_size = 1)

### 2.1.3. SphereNet3D Family

In this case, we used the fundamental base architecture of SphereNet3D, which is similar to the 3D O-Net, instead consisting of four sets of encoder-decoder pairs.

## 2.2. Loss Functions

In our segmentation pipeline, we have employed four different loss functions coupled with aforementioned suitable models.

### 2.2.1. Cross-entropy and dice loss

We used a combination of cross entropy and dice loss function where the dice loss is taken as the average of four labels including the background.

$$Loss = cross\_entropy\_loss + \frac{1}{4} \sum_{i=1}^{4} (dice\_loss_i) \qquad \ldots(1)$$

### 2.2.2. BasnetHybrid Loss

We used BasnetHybrid loss which is a combination of cross-entropy loss, multi-scale structural similarity loss, and Jaccard index loss [11].

$$Loss = cross\_entropy\_loss + \frac{1}{4} \sum_{i=1}^{4} (ms\_ssim\_loss_i + jaccard\_loss_i) \qquad \ldots(2)$$

The ms_ssim_loss$_i$ and jaccard_loss$_i$ are computed individually for all four classes and then averaged.

### 2.2.3. Blob loss with BasnetHybrid loss

We used a combination of Blob loss along with BasnetHybrid loss to create a tailored loss function for this particular task [12].

$$Loss = \alpha * BasnetHybridLoss_{global} + \beta * BasnetHybridLoss_{blob} \qquad \ldots(3)$$

where $BasnetHybridLoss_{global}$ is the loss function defined in 2.2.2, $BasnetHybridLoss_{blob}$ is the loss function average loss across all connected components of the ground-truth segmentation mask. It is defined as below:

$$BasnetHybridLoss_{blob} = \frac{1}{4*n} \sum_{i=1}^{4} \sum_{n=1}^{k} basnet\_hybrid\_loss(g_n^i, p_n^i) \qquad \ldots(4)$$

where $g_n^i, p_n^i$ are ground-truth mask and predicted mask for the nth component of ith class of an image.

## 2.3. Ensemble Method

We used a tailored ensemble approach using the softmax probabilities for each class which is then averaged over all models to produce the final prediction. If we denote the softmax probability of class $c$ from model $m$ as $P_{m,c}$, the final probability for class c is given by:

$$P_c = \frac{1}{M} \sum_{m=1}^{M} P_{m,c} \qquad \ldots(5)$$

where $M$ is the total number of models in the ensemble.

This ensemble approach enhances the robustness and accuracy of the final segmentation results by aggregating information from different models, potentially leading to more precise predictions.

## 2.4. Post Processing

Post-processing techniques are applied to the combined predictions to refine segmentation quality:

1. Size Filtering Based on Voxel Volumes: This technique removes small isolated regions, eliminating noise.
2. Morphological Reconstruction: A more advanced method for interpolation of voxels violating constraints and smoothing boundaries for all labels.

# 3. Results

## 3.1. Evaluation Metrics

### 3.1.1. Dice Score and Hausdorff distance

The models submitted to the BraTS challenge are assessed utilizing two key metrics: the lesion_wise Dice score and the 95th percentile lesion_wise Hausdorff distance. These metrics are applied across the whole tumor, core tumor, and active tumor sub-regions. On a fundamental level, the Dice score quantifies the overlap between the predicted segmentations and the actual ground truth, while the Hausdorff distance calculates the extent to which the predicted and ground truth segmentations deviate from one another. While in the case of lesion_wise, the dice scores and HD95 scores are calculated for each lesion (or component) individually and then it penalizes all the False Positives and the False Negatives with a 0 score for dice and 374 for HD95, after which the mean is taken for the particular caseID.

## 3.2. Quantitative Performance

Table - 1: Comparison of different evaluation metrics for enhancing tumor

| Models | Lesion_wise Dice | Dice | Lesion_wise Hausdorff95 | Hausdorff95 |
|---|---|---|---|---|
| Ensemble_Cross Entropy and Dice Loss | 0.50 | 0.82 | 162.07 | **9.14** |
| Ensemble_BasnetHybrid Loss | 0.59 | 0.63 | **53.74** | 35.62 |
| Ensemble_Blob loss | **0.61** | **0.82** | 113.68 | 32.60 |

Table - 2: Comparison of different evaluation metrics for tumor core

| Models | Lesion_wise Dice | Dice | Lesion_wise Hausdorff95 | Hausdorff95 |
|---|---|---|---|---|
| Ensemble_Cross Entropy and Dice Loss | 0.49 | 0.82 | 163.05 | **9.16** |
| Ensemble_BasnetHybrid Loss | **0.64** | 0.70 | **50.61** | 32.35 |
| Ensemble_Blob | 0.62 | **0.82** | 114.31 | 31.59 |
| loss | | | | |

Table - 3: Comparison of different evaluation metrics for whole tumor

| Models | Lesion_wise Dice | Dice | Lesion_wise Hausdorff95 | Hausdorff95 |
|---|---|---|---|---|
| Ensemble_Cross Entropy and Dice Loss | 0.61 | 0.86 | 130.86 | **6.18** |
| Ensemble_BasnetHybrid Loss | **0.75** | 0.82 | **59.83** | 31.47 |
| Ensemble_Blob loss | 0.73 | **0.87** | 81.28 | 28.10 |

In our assessment of various models, we submitted the results to the synapse portal for testing the results on the validation. The findings revealed that the ensemble training approach with specific loss demonstrates better effectiveness in comparison to other loss functions for particular cases or segmentation labels.

Furthermore, an individual evaluation of predictions indicated that the ensemble training approach with the same loss function might surpass other labels as outlined in Table 1,2,3. This indicates that there can be an ensemble model along with a loss function which might be robust and be generalizable for a larger dataset. However for specific types of segmentation and task in hand, the approaches might differ.

## 3.3. Visual Comparison

Figure - 3 presents a comparative analysis of prediction results from two distinct approaches for models: single model training and the ensemble method.

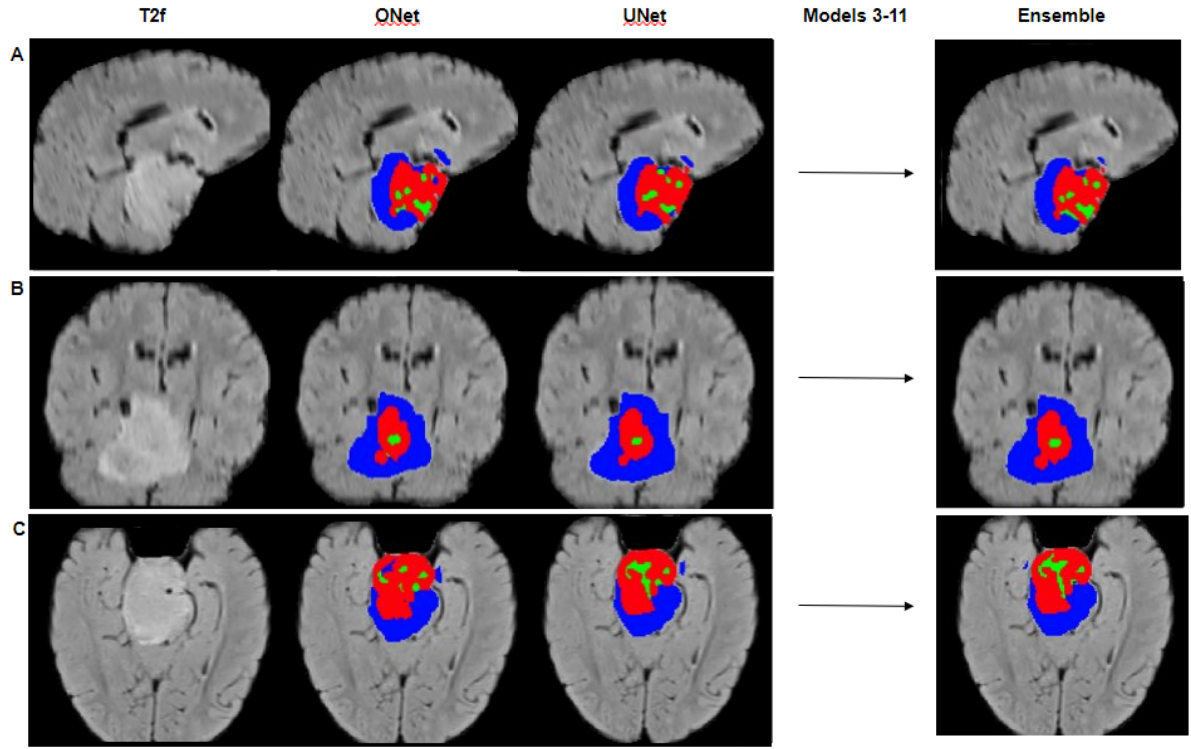

Figure - 3. A, B, and C depict three slice planes(sagittal, coronal, axial) from the same CaseID along with the predictions generated by the UNet3D, ONet3D and the ensemble mode respectively.

It is evident that single model training appears to capture fewer tumor and enhanced tumor regions in the models' predictions. This is effectively addressed when utilizing the ensemble training approach.

## 4. Discussion and Conclusion

In this paper, we present the significance of employing an ensemble strategy and various loss functions. We have trained eleven different models to train on the dataset and subsequently combined their results. Due to time constraints, it has been challenging to precisely discern which models excel in specific regions of the brain, thereby underlining the complexity and distinction involved in this analysis. However, we observed that there is an increase in the evaluation metrics in the ensemble model than the single models which emphasizes the fact that combining different architectures can help us precisely perform segmentation in the brain or other medical images. Future direction will be focused on identifying brain region-specific model behavior, to create a significantly improved ensemble, alongside continuing base model training for a greater number of epochs.